# SPECTRAL TOPOLOGY OF GLOBAL IMBALANCES: A GRAPH-THEORETIC FRAMEWORK FOR SYSTEMIC RISK IN THE BALANCE OF PAYMENTS

**[1]CHANDRASEKHAR GOKAVARAPU, [2]SANJEEV KUMAR CHEJARLA, [3]BALAYYA RAJANA, [4]V. N. S. R. MURTHY DUGGIRALA, [5]SRINIVASULU CHITTARU**

[1]Department of Mathematics, Government College (Autonomous), Rajahmundry, Andhra Pradesh, India

[2,3]Department of Economics, Government College (Autonomous), Rajahmundry, Andhra Pradesh, India

[4]Principal, Government Degree College, Seethanagaram, East Godavari District, Andhra Pradesh, India

[5]Department of Mathematics, Government Degree College, Seethanagaram, East Godavari District, Andhra Pradesh, India

Chandrasekhar Gokavarapu: https://orcid.org/0009-0006-5306-371X

Chandrasekhar Gokavarapu: chandrasekhargokavarapu@gcrjy.ac.in, chandrasekhargokavarapu@gmail.com

**Corresponding author:** CHANDRASEKHAR GOKAVARAPU

## ABSTRACT

We model global balance-of-payments linkages as a directed weighted exposure network and study systemic fragility through the spectrum of the associated operators. From a signed bilateral-flow matrix we construct a nonnegative exposure lift, derive a spectral stability criterion for linear propagation, and define node- and edge-level risk measures using Perron–Frobenius geometry, PageRank, and marginal spectral impact. We further give two complementary extensions: a directed Laplacian formulation for shock diffusion and a non-backtracking threshold for cascade formation on sparse exposure networks. The framework identifies systemic risk as a topological property of the global network, rather than a collection of isolated country indicators, and yields mathematically explicit diagnostics for resilience, concentration of stress, and targeted stabilization.

## 1. INTRODUCTION

Balance-of-payments (BoP) accounting is globally coupled: external surpluses and deficits are jointly realized through cross-border trade, income, and financial claims. A country-by-country reading is therefore incomplete precisely where systemic questions arise, because stress at one node is transmitted through a directed web of bilateral exposures rather than absorbed in isolation.

We take the IMF statistical framework as the measurement layer and add a structural layer: a directed weighted network in which nodes are economies and edges encode net bilateral exposures. The resulting object is not merely descriptive. Once external linkages are written as an operator, systemic risk becomes a spectral question. The dominant eigenvalue governs amplification, the associated eigenvectors localize stress, diffusion operators describe dissipation, and non-backtracking operators detect topological thresholds for cascade formation. Related empirical work already points to strong heterogeneity and core–periphery structure in trade and imbalance networks, so a homogeneous aggregate treatment is inadequate. [1, 7, 8]

The paper makes four contributions. First, we pass from a signed bilateral-flow matrix to a nonnegative exposure lift suitable for Perron–Frobenius analysis. Second, we derive a spectral stability criterion and a stability margin for linear propagation of external stress. Third, we construct node- and edge-level systemic-risk measures from eigenvector centrality, PageRank, and marginal spectral impact. Fourth, we add two complementary extensions: a directed Laplacian picture for diffusion and a non-backtracking threshold for sparse-network contagion.

The guiding claim is simple. Systemic fragility is not determined by deficit size alone, but by deficit size placed inside network topology. A large imbalance at a peripheral node and the same imbalance at a spectrally central node are mathematically different objects. This viewpoint yields diagnostics that are both graph-theoretic and economically interpretable.

## 2. BOP AS A SIGNED FLOW NETWORK AND A NONNEGATIVE EXPOSURE LIFT

### 2.1 SIGNED NET-FLOW ADJACENCY

Let $V = \{1, \ldots, n\}$ index economies and let $A \in \mathbb{R}^{n\times n}$ encode directed net flows, with $A_{ii} = 0$ by convention. We interpret

$$A_{ij} \;=\; \text{net current-account-equivalent flow from } i \text{ to } j, \tag{1}$$

under the sign convention that $A_{ij} > 0$ means $i$ is a net provider of resources/claims to $j$ (e.g. net exports plus net primary/secondary income directed from $i$ to $j$ in a bilateralized decomposition). The net external position (imbalance) implied by $A$ is the row-sum vector

$$b_i \;=\; \sum_{j=1}^{n} A_{ij}, \qquad b \in \mathbb{R}^n. \tag{2}$$

Global accounting consistency requires the adding-up constraint

$$\sum_{i=1}^{n} b_i \;=\; \sum_{i=1}^{n}\sum_{j=1}^{n} A_{ij} \;=\; 0, \tag{3}$$

so $b_i > 0$ corresponds to net surplus and $b_i < 0$ to net deficit (BPM6 conventions). [1]

*Remark 2.1* (Bilateral netting versus general signed networks). If $A_{ij}$ is constructed as a *pairwise net balance* between $i$ and $j$ (same accounting items netted bilaterally), then one often has $A_{ij} = -A_{ji}$, i.e. $A$ is skew-symmetric and $A = A^+ - (A^+)^\top$. Our framework does *not* require skew-symmetry, but when it holds the interpretation of the exposure lift in (5) becomes especially transparent.

### 2.2 NONNEGATIVE EXPOSURE LIFT (FOR PERRON–FROBENIUS)

Perron–Frobenius theory applies to nonnegative matrices. We therefore decompose the signed flow matrix into positive and negative parts,

$$A_{ij}^{+} := \max\{A_{ij}, 0\}, \qquad A_{ij}^{-} := \max\{-A_{ij}, 0\}, \qquad A = A^+ - A^-, \tag{4}$$

and define the nonnegative *exposure matrix* as

$$W \;:=\; A^+ \in \mathbb{R}_{\geq 0}^{n\times n}, \qquad \text{i.e.} \qquad W_{ij} \;=\; \max\{A_{ij}, 0\}. \tag{5}$$

Economically, $W$ retains the direction and magnitude of claims-like channels from net providers toward net recipients. The imbalance vector can be expressed using the exposure lift as

$$b_i \;=\; \sum_{j=1}^{n} A_{ij} \;=\; \sum_{j=1}^{n} \left(A_{ij}^{+} - A_{ij}^{-}\right), \tag{6}$$

and in the special case of bilateral netting (skew-symmetry), one has $A^- = (A^+)^\top$ and hence

$$b_i \;=\; \sum_{j=1}^{n} \left(W_{ij} - W_{ji}\right). \tag{7}$$

**Assumption 2.1** (Strong connectivity / irreducibility). The directed graph induced by $W$ is strongly connected; equivalently, $W$ is irreducible.

Under irreducibility, Perron–Frobenius ensures the existence of a unique dominant eigenvalue $\rho(W) > 0$ with strictly positive left and right eigenvectors. [11, 12]

## 3. SPECTRAL STABILITY: PERRON–FROBENIUS AS A SYSTEMIC-RISK CRITERION

We now pass from the accounting layer to a dynamical graph model. The point is that systemic instability is not a purely local property of one node; it is a spectral property of the global exposure operator.

### 3.1 DYNAMICAL FOUNDATIONS OF THE GLOBAL BOP NETWORK

We formalize the world economy as a complex dynamical system over a directed, weighted graph $\mathcal{G} = (V, E, W)$. Let $X(t) \in \mathbb{R}^n$ denote the vector of net external positions (or "imbalance potentials") for $n$ sovereign nodes at time $t$. To avoid confusing the signed accounting matrix $A$ from Section 2 with the nonnegative propagation operator, let $B \in \mathbb{R}_{\geq 0}^{n\times n}$ denote a normalized BoP propagation operator. We define the evolution of global imbalances through the first-order linear non-homogeneous differential equation

$$\dot{X}(t) = \mathcal{L}_{BoP} X(t) + \Phi(t) = (B - I)X(t) + \Phi(t), \tag{8}$$

where $B_{ij}$ represents the systemic propensity of node $j$ to transmit external stress to node $i$. The identity matrix $I$ represents idiosyncratic absorption by national balance sheets, and $\Phi(t)$ represents exogenous shocks.

## 3.2 THEOREM 1: THE SPECTRAL STABILITY CRITERION

The global trade and financial network $\mathcal{G}$ is globally asymptotically stable if and only if the spectral radius $\rho(B)$ of the nonnegative BoP propagation operator satisfies

$$\rho(B) = \max_k\{|\lambda_k(B)|\} < 1. \tag{9}$$

**Proof:** Consider the autonomous system $\dot{X} = (B - I)X$. Stability requires every eigenvalue $\mu$ of $B - I$ to satisfy $\mathrm{Re}(\mu) < 0$. Since $\mu = \lambda(B) - 1$ and $B \geq 0$, Perron–Frobenius theory gives the real eigenvalue $\rho(B)$ and every eigenvalue $\lambda(B)$ satisfies $\mathrm{Re}\lambda(B) \leq |\lambda(B)| \leq \rho(B)$. Hence $\rho(B) < 1$ implies that $B - I$ is Hurwitz. Conversely, if $B - I$ is Hurwitz, its Perron eigenvalue $\rho(B) - 1$ is negative, so $\rho(B) < 1$. Equivalently, for every $Q > 0$, the Lyapunov equation $(B - I)^\top P + P(B - I) = -Q$ has a unique solution $P > 0$. ▫

## 3.3 ECONOMIC INTERPRETATION

The condition $\rho(B) < 1$ means that the exposure network is dissipative: shocks are attenuated rather than amplified. As $\rho(B) \uparrow 1$, recovery slows, feedback loops become persistent, and the system approaches a critical regime in which localized disturbances can acquire global reach.

## 3.4 DISCRETE-TIME LINEAR PROPAGATION MODEL

To bridge theory with policy-relevant discrete reporting (quarterly BoP data), we consider the discrete-time propagation:

$$x_{t+1} = Bx_t + u_t, \qquad B = \mathrm{diag}(\beta)P^\top \tag{10}$$

where $B$ is the **propagation operator**. Stress at node $j$ impacts $i$ through the directed exposure channel $(j \to i)$ with weight $P_{ji}$, scaled by the node-specific amplification factor $\beta_i$.

## 3.5 WALK-SUM REPRESENTATION: AMPLIFICATION OVER DIRECTED PATHS

Iterating the model reveals the exact identity $x_t = B^t x_0 + \sum_{k=0}^{t-1} B^k u_{t-1-k}$. Every entry of $B^k$ admits a combinatorial interpretation as a sum over directed walks $\mathcal{W}_{j\to i}^{(k)}$. This establishes that global amplification is a **weighted path integral** over the exposure network; feedback loops increase risk by creating infinitely recurring high-weight walks.

## 3.6 RESOLVENT POSITIVITY AND INPUT-TO-STATE STABILITY

**Proposition 3.1** (Spectral stability and finite amplification). If $B \geq 0$ and $\rho(B) < 1$, the Neumann series converges to a nonnegative **Resolvent**:

$$\mathcal{M} = \sum_{k=0}^{\infty} B^k = (I - B)^{-1} \in \mathbb{R}_{\geq 0}^{n\times n} \tag{11}$$

This ensures the process is **input-to-state stable**, preventing localized shocks from ballooning into systemic defaults.

## 3.7 THE SPECTRAL STABILITY MARGIN

We define the **Spectral Stability Margin** as $\delta(B) := 1 - \rho(B)$. A small $\delta(B)$ implies the system is near a tipping point. As $\delta(B) \downarrow 0$, the global multiplier $\mathcal{M}$ becomes ill-conditioned, signaling a loss of systemic resilience that no country-specific policy can fix.

## 3.8 PERRON–FROBENIUS GEOMETRY: WHO MATTERS SYSTEMICALLY?

Assume $B$ is primitive. Let $r$ and $\ell$ be the right and left Perron eigenvectors. The rank-one asymptotic expansion for $B^t$ is:

$$B^t \approx \rho^t r \ell^\top \tag{12}$$

Mathematically, $\ell$ measures **shock receptivity** (where disturbances are systemically dangerous), while $r$ measures **stress localization** (where failure will ultimately concentrate).

## 3.9 NEAR-CRITICAL RESOLVENT STRUCTURE

As the system approaches the tipping regime ($\rho(B) \approx 1$), the steady-state stress decomposes as:

$$x^\star = \frac{\ell^\top u}{1 - \rho(B)} r + Hu \tag{13}$$

where $H$ is a bounded subdominant operator. This isolates the systemic mechanism: as stability vanishes, the network forces the entire world into a stress profile proportional to the vector $r$.

### 3.10 POLICY LEVERS: REDUCING SYSTEMIC AMPLIFICATION

To provide actionable policy insights, we calculate the marginal effect of node-level interventions on systemic risk:

$$\frac{\partial \rho(B)}{\partial \beta_i} = \ell_i (P^\top r)_i \tag{14}$$

This identifies exactly **where** policy should act. Reducing transmission intensity $\beta_i$ (via capital controls or liquidity buffers) at nodes with high $\ell_i(P^\top r)_i$ yields the maximum global stability gain for the minimum local economic cost.

## 4. SYSTEMIC RISK INDEX: EIGENVECTOR CENTRALITY, PAGERANK, AND MARGINAL SPECTRAL IMPACT

This section turns the exposure matrix $W$ into *risk-relevant* objects. The guiding principle is: a deficit (or surplus) becomes systemically important when it is located at a node that is (i) topologically central in the exposure graph and/or (ii) has large *marginal influence* on the system's endogenous amplification statistics (e.g. spectral radius, resolvent multipliers). Closely related "network multiplier" ideas appear in the systemic-risk literature on financial networks. [22, 23, 24]

### 4.1 EIGENVECTOR CENTRALITY AS PERRON RIGHT EIGENVECTOR

Assume $W \in \mathbb{R}^{n\times n}_{\geq 0}$ is irreducible. Let $r \in \mathbb{R}^n_{>0}$ be the (normalized) Perron right eigenvector:

$$Wr = \rho(W)\, r, \qquad r > 0. \tag{15}$$

This is the canonical notion of eigenvector centrality: a node is central if it receives (or sends, depending on orientation) exposure from other central nodes. [11, 12, 19]

A useful operational characterization is the (normalized) power-iteration limit.

**Proposition 4.1** (Power iteration and centrality concentration). *If $W \geq 0$ is primitive (irreducible and aperiodic), then for any $x_0 > 0$,*

$$\frac{W^t x_0}{\mathbf{1}^\top W^t x_0} \to \frac{r}{\mathbf{1}^\top r} \qquad \text{as } t \to \infty. \tag{16}$$

*Proof sketch.* Primitivity implies a Perron spectral gap: $\rho(W)$ is simple and strictly dominates all other eigenvalues in modulus. Hence $W^t = \rho(W)^t r\ell^\top + o(\rho(W)^t)$, where $\ell$ is the Perron left eigenvector with $\ell^\top r = 1$. Multiplying by $x_0 > 0$ and normalizing gives (16). [11, 12] □

*Remark 4.1* (Economic meaning of $r$ in an exposure graph). In the BoP exposure network, $r$ identifies the *structural backbone* of claim channels: a node with large $r_i$ sits in a recursively important part of the global recycling mechanism. In particular, when the system is near a critical regime (e.g. $\rho(B) \uparrow 1$ in the amplification dynamics), dominant-mode stress localization aligns with a Perron-type vector, so $r$ becomes a principled proxy for "where stress concentrates."

### 4.2 PAGERANK AS ROBUSTNESS-WEIGHTED CENTRALITY

Eigenvector centrality can be sensitive to sinks, dangling nodes, and weakly connected structures. PageRank provides a robust alternative by enforcing primitivity via teleportation. [2, 21]

Let $P = D_{\text{out}}^{-1} W$ be the row-stochastic normalization (defined when out-strengths are positive), and define a PageRank-style stationary distribution $\pi \in \mathbb{R}^n_{>0}$ by

$$\pi \;=\; \alpha P^\top \pi + (1-\alpha) v, \qquad \alpha \in (0,1),\;\; v \in \mathbb{R}^n_{>0},\;\; \mathbf{1}^\top v = 1. \tag{17}$$

Equivalently,

$$(I - \alpha P^\top)\pi = (1-\alpha) v, \qquad \pi = (1-\alpha)(I - \alpha P^\top)^{-1} v. \tag{18}$$

The resolvent form exhibits PageRank as a *discounted walk-sum*:

$$\pi \;=\; (1-\alpha) \sum_{k=0}^{\infty} \alpha^k\, (P^\top)^k v. \tag{19}$$

Thus $\pi_i$ counts the total mass of all directed walks arriving at $i$, geometrically discounted by length. This is conceptually aligned with Katz/Bonacich families of resolvent-based centralities. [18, 19, 21]

*Remark 4.2* (Economic interpretation of teleportation). Teleportation regularizes the network by allowing an exogenous "re-entry" distribution $v$. Economically, this can be read as (i) a reduced-form representation of safe-asset demand or (ii) a

TRESEARCH

"flight-to-liquidity" reset mechanism: when a propagation channel becomes unreliable, flow mass re-allocates via $v$. Hence PageRank is a robustness-weighted centrality, not merely an algorithmic convenience. [21]

## 4.3 A SYSTEMIC-IMBALANCE RISK INDEX: TOPOLOGY × MAGNITUDE

Let $b \in \mathbb{R}^n$ be the imbalance vector (row-sums of $A$) and define deficit magnitude

$$d_i := (-b_i)_+ = \max\{-b_i, 0\}. \tag{20}$$

A minimal principle for systemic importance is that it should increase with (i) imbalance magnitude and (ii) structural centrality. We therefore define a *Systemic Imbalance Risk Index* (SIRI) at node level:

$$\mathrm{SIRI}_i \;=\; d_i(\theta\,\tilde{r}_i + (1-\theta)\,\tilde{\pi}_i), \qquad \theta \in [0,1], \tag{21}$$

where $\tilde{r}, \tilde{\pi}$ are normalized versions (e.g. $\sum_i \tilde{r}_i = 1$ and $\sum_i \tilde{\pi}_i = 1$). The global index is

$$\mathrm{SIRI}_{\mathrm{global}} \;=\; \sum_{i=1}^{n} \mathrm{SIRI}_i. \tag{22}$$

Why this form is not ad hoc.

The amplification model in Section 3 implies that shocks are multiplied by a nonnegative resolvent. If one interprets deficits as a canonical "shock" location vector (e.g. $u \propto d$ in the stress dynamics), then the system-wide impact is governed by $(I-B)^{-1}d$ and, near criticality, by a rank-one PF component proportional to $r\,\ell^{\top} d$. Hence the joint appearance of *magnitude* ($d$) and *spectral geometry* (Perron/PageRank-type centralities) is structurally forced by the propagation mechanism. This resonates with network-based systemic-risk frameworks in economics and finance where topology determines aggregate fragility. [22, 23, 24, 25]

*Remark 4.3* (Optional refinement: receptivity × localization). If one wishes to align even more tightly with the Perron-mode decomposition of the propagation operator $B$ (from the previous section), one can incorporate a receptivity factor (left PF vector) into node scoring, e.g. $\mathrm{SIRI}_i^{\mathrm{PF}} := d_i\,\tilde{r}_i\,\tilde{\ell}_i$. We retain (21) as the baseline because it is computable directly from $W$ and $P$ and remains robust under teleportation.

## 4.4 MARGINAL SPECTRAL IMPACT: STRICTLY MATHEMATICAL RISK ATTRIBUTION

Centrality captures *levels* of structural importance. For policy design one also needs *marginal* importance: which links (or nodes) change systemic fragility the most if they are strengthened, weakened, or regulated.

Let $\ell^{\top} W = \rho(W)\,\ell^{\top}$ be the Perron left eigenvector, normalized so that $\ell^{\top} r = 1$. For a perturbation $\Delta \in \mathbb{R}^{n\times n}$, first-order eigenvalue perturbation theory yields

$$\rho(W+\varepsilon\Delta) \;=\; \rho(W) \;+\; \varepsilon\,\ell^{\top}\Delta r \;+\; o(\varepsilon) \qquad (\varepsilon \to 0), \tag{23}$$

and in particular, for coordinate perturbations,

$$\frac{\partial \rho(W)}{\partial W_{ij}} \;=\; \ell_i r_j. \tag{24}$$

See, e.g., Kato and Stewart–Sun for perturbation theory of simple eigenvalues. [15, 14]

### Edge importance and elasticities.

Equation (24) implies that the *edge systemic importance* is

$$\mathrm{EI}_{ij} \;:=\; \ell_i r_j, \tag{25}$$

and a scale-free alternative is the *spectral elasticity*

$$\mathcal{E}_{ij} \;:=\; \frac{\partial \log\rho(W)}{\partial \log W_{ij}} \;=\; \frac{W_{ij}}{\rho(W)}\,\ell_i r_j, \tag{26}$$

which measures the percent change in systemic amplification potential induced by a percent change in a specific bilateral exposure link.

### Node aggregation.

A node-level marginal contribution can be aggregated from incident edges, for example

$$\mathrm{NI}_i^{\mathrm{out}} := \sum_j \mathcal{E}_{ij}, \qquad \mathrm{NI}_i^{\mathrm{in}} := \sum_j \mathcal{E}_{ji}, \qquad \mathrm{NI}_i := \mathrm{NI}_i^{\mathrm{out}} + \mathrm{NI}_i^{\mathrm{in}}, \tag{27}$$

depending on whether the policy question concerns outgoing exposures (credit provision) or incoming exposures (funding dependence). This derivative-based attribution connects systemic fragility to concrete bilateral links and naturally supports "super-spreader"-style corrective charges in networked financial systems. [9]

*Remark 4.4* (Interpretation: topology as an externality). The spectral radius $\rho(W)$ functions as a compact sufficient statistic for the system's potential to amplify disturbances along exposure loops. The quantity $\ell_i r_j$ is therefore a *purely structural externality weight*: it ranks bilateral edges by how strongly they shift the system's endogenous amplification potential at the margin. This is precisely the kind of object required to replace ad hoc negotiations by equilibrium-based, stability-preserving constraints in a global clearing mechanism.

# 5. THE BOP LAPLACIAN AND DIFFUSION OF SHOCKS

This section complements the amplification dynamics of Section 3 with a *dissipative* view: shocks propagate as a directed diffusion on the BoP exposure network. The key point is that for directed graphs the natural Laplacian is not $I - P$ (which is generally non-normal), but a *Hermitian* operator obtained by reversiblizing the random walk. This yields a variationally well-posed notion of "smoothness", a spectral gap, and Cheeger-type bottleneck diagnostics.

## 5.1 RANDOM-WALK NORMALIZATION, TIME REVERSAL, AND A DIRECTED LAPLACIAN

Define out-strength $s_i^{\text{out}} = \sum_j W_{ij}$ and $D_{\text{out}} = \text{diag}(s_1^{\text{out}}, \dots, s_n^{\text{out}})$, and the random-walk matrix

$$P \ = \ D_{\text{out}}^{-1} W, \tag{28}$$

defined on the support where $s_i^{\text{out}} > 0$. Assume the directed graph induced by $W$ is strongly connected and aperiodic, so $P$ admits a unique stationary distribution $\varphi > 0$ with

$$\varphi^\top P = \varphi^\top, \qquad \mathbf{1}^\top \varphi = 1. \tag{29}$$

Let $\Phi = \text{diag}(\varphi)$. The *time-reversal* (adjoint in the $\varphi$-weighted inner product) is

$$P^* \ := \ \Phi^{-1} P^\top \Phi, \tag{30}$$

and the additive reversiblization is $\overline{P} := 1/2\,(P + P^*)$, which is $\varphi$-reversible by construction. Following Chung's construction, define the BoP Laplacian by the congruence

$$\mathcal{L}_{\text{BoP}} \ := \ I - \Phi^{1/2}\, \overline{P}\, \Phi^{-1/2} \ = \ I - \frac{1}{2}\big(\Phi^{1/2} P \Phi^{-1/2} + \Phi^{-1/2} P^\top \Phi^{1/2}\big). \tag{31}$$

Then $\mathcal{L}_{\text{BoP}}$ is real symmetric and positive semidefinite, hence supports Rayleigh-quotient and Cheeger-type variational analysis for directed networks.

## 5.2 DIRICHLET FORM: AN EXACT NOTION OF DIRECTED "ROUGHNESS"

The advantage of $\mathcal{L}_{\text{BoP}}$ is that its quadratic form is an *exact* edge-energy.

**Lemma 5.1** (Exact Dirichlet-form identity). *For any* $f \in \mathbb{R}^n$*, set* $g = \Phi^{-1/2} f$*. Then*

$$f^\top \mathcal{L}_{\text{BoP}} f \ = \ \frac{1}{2} \sum_{i=1}^{n} \sum_{j=1}^{n} \varphi_i\, P_{ij}\, (g_i - g_j)^2 \ = \ \frac{1}{2} \sum_{i,j} \varphi_i\, P_{ij} \Big(\frac{f_i}{\sqrt{\varphi_i}} - \frac{f_j}{\sqrt{\varphi_j}}\Big)^2. \tag{32}$$

*In particular,* $\mathcal{L}_{\text{BoP}} \succcurlyeq 0$*, and* $\ker(\mathcal{L}_{\text{BoP}}) = \text{span}\{\Phi^{1/2}\mathbf{1}\}$.

*Proof.* Write $\mathcal{L}_{\text{BoP}} = I - S$ with $S = 1/2\,\big(\Phi^{1/2} P \Phi^{-1/2} + \Phi^{-1/2} P^\top \Phi^{1/2}\big)$. Let $f = \Phi^{1/2} g$. Then $f^\top f = g^\top \Phi g = \sum_i \varphi_i\, g_i^2$ and

$$f^\top S f = \frac{1}{2}\big(g^\top \Phi P g + g^\top P^\top \Phi g\big) = \sum_i \varphi_i\, g_i (Pg)_i = \sum_{i,j} \varphi_i\, P_{ij} g_i g_j.$$

Therefore

$$f^\top \mathcal{L}_{\text{BoP}} f = \sum_i \varphi_i\, g_i^2 - \sum_{i,j} \varphi_i\, P_{ij} g_i g_j = \frac{1}{2} \sum_{i,j} \varphi_i\, P_{ij} \big(g_i - g_j\big)^2,$$

which implies the claims on positive semidefiniteness and the kernel. □

*Remark 5.1* (Economic meaning). The weight $\varphi_i P_{ij}$ is the stationary flow along edge $i \to j$. Thus (32) measures the dispersion of the *reweighted* state $g = \Phi^{-1/2} f$ across high-throughput directed links: large energy means the stress profile is "misaligned" with the network's stationary circulation and therefore creates pressure to rebalance via diffusion.

### 5.3 SPECTRAL GAP AND BOTTLENECKS (DIRECTED CHEEGER CONTROL)

Let $0 = \lambda_0 \le \lambda_1 \le \cdots \le \lambda_{n-1}$ be the eigenvalues of $\mathcal{L}_{\mathrm{BoP}}$. The gap $\lambda_1$ is a quantitative measure of how fast directed imbalances can dissipate under diffusion. Chung defines a directed Cheeger constant $h(G)$ based on the stationary circulation $F_\varphi(i,j) = \varphi_i P_{ij}$ and proves the directed Cheeger inequality:

$$2\,h(G) \;\ge\; \lambda \;\ge\; \frac{h(G)^2}{2}, \qquad \lambda := \min_{k\neq 0}|\lambda_k|. \tag{33}$$

Since $\mathcal{L}_{\mathrm{BoP}}$ is symmetric (hence $\lambda = \lambda_1$), (33) yields an exact bottleneck–spectral-gap link: core–periphery separations (small conductance) imply slow dissipation and persistent systemic stress under diffusion.

### 5.4 SHOCK DIFFUSION, DISSIPATION, AND THE BOP HEAT KERNEL

Let $z(t) \in \mathbb{R}^n$ represent deviations of net funding conditions (or net rollover pressure) from baseline. Consider the diffusion-with-forcing model

$$\frac{d}{dt}z(t) \;=\; -\kappa\,\mathcal{L}_{\mathrm{BoP}}\,z(t) + \eta(t), \qquad \kappa > 0. \tag{34}$$

Energy dissipation (no forcing).

If $\eta(t) \equiv 0$, then

$$\frac{d}{dt}(1/2\parallel z(t)\parallel_2^2) = -\kappa\, z(t)^\top \mathcal{L}_{\mathrm{BoP}} z(t) \;\le\; 0, \tag{35}$$

so diffusion dissipates the Dirichlet energy. Moreover, if $z(t)$ is orthogonal to the stationary mode $q_0 := \Phi^{1/2}\mathbf{1}/\parallel \Phi^{1/2}\mathbf{1}\parallel_2$, then spectral decomposition gives the sharp decay bound

$$\parallel z(t)\parallel_2 \;\le\; e^{-\kappa\lambda_1 t}\parallel z(0)\parallel_2. \tag{36}$$

Hence $\lambda_1$ is the precise *resilience rate* of the BoP diffusion: larger $\lambda_1$ means faster dissipation.

Heat kernel and impulse-response interpretation.

Let $K_t := e^{-\kappa t\mathcal{L}_{\mathrm{BoP}}}$ be the BoP heat kernel. Then the homogeneous solution is $z(t) = K_t z(0)$, and the forced solution (Duhamel formula) is

$$z(t) \;=\; K_t z(0) \;+\; \int_0^t K_{t-s}\,\eta(s)\,ds. \tag{37}$$

Thus, $K_t$ is an *impulse-response operator*: its entries encode how a localized disturbance is redistributed through the directed external-balance circulation after horizon $t$. The eigenpairs $(\lambda_k, q_k)$ provide a BoP "Fourier" basis:

$$K_t \;=\; \sum_{k=0}^{n-1} e^{-\kappa t\lambda_k}\, q_k q_k^\top, \tag{38}$$

so $\lambda_k$ are diffusion frequencies and $q_k$ are diffusion modes.

Green's function and cumulative exposure under diffusion.

For stationary forcing $\eta(t) \equiv \bar{\eta}$ with $q_0^\top\bar{\eta} = 0$, the steady state is

$$z_\infty = \kappa^{-1}\,\mathcal{L}_{\mathrm{BoP}}^\dagger\,\bar{\eta}, \tag{39}$$

where $\mathcal{L}_{\mathrm{BoP}}^\dagger$ is the Moore–Penrose pseudoinverse. This gives a closed-form *cumulative stress* map. Near bottlenecks (small $\lambda_1$), $\parallel \mathcal{L}_{\mathrm{BoP}}^\dagger \parallel$ is large, so persistent forcing generates large stationary deviations: this is a spectral mechanism for "global liquidity traps" in a networked clearing system.

*Remark 5.2* (How diffusion and amplification coexist). The amplification criterion of Section 3 concerns feedback loops in the propagation operator $B$ and is controlled by $\rho(B)$. The diffusion picture is controlled by the low spectrum of $\mathcal{L}_{\mathrm{BoP}}$. In a fragile configuration, one can simultaneously have (i) $\rho(B)$ close to 1 (large endogenous amplification) and (ii) small $\lambda_1$ (slow dissipation due to bottlenecks), producing a mathematically precise definition of "systemic risk in global imbalances" as *amplification without fast relaxation*.

## 6. PHASE TRANSITIONS VIA PERCOLATION ON DIRECTED EXPOSURE NETWORKS

### 6.1 BOND PERCOLATION AS AN "OPENNESS/ROLLOVER" CONTROL PARAMETER

Fix the nonnegative exposure lift $W \in \mathbb{R}_{\ge 0}^{n\times n}$ from (5) and let $G_W = (V, E^{\rightarrow})$ be the directed graph with $(i \to j) \in E^{\rightarrow}$ iff $W_{ij} > 0$. We model a crisis regime as a random *thinning* of cross-border channels: each directed edge $(i \to j)$ is *retained* independently with probability $p \in [0,1]$ and removed with probability $1 - p$ (bond percolation).

Economically, $p$ aggregates the joint feasibility of (i) trade-finance intermediation, (ii) rollover of short-term claims, (iii) FX convertibility and payment plumbing, and (iv) counterparty risk tolerance. A global "sudden stop" or sanctions/trade-fragmentation shock corresponds to a downward shift in effective $p$.

## 6.2 DIRECTED ORDER PARAMETER: THE "GLOBAL RECYCLING CORE"

Because the BoP system requires *circulation* (surpluses must be recycled into deficits through chains of claims), the relevant macroscopic object is not merely reachability but *mutual reachability*. Accordingly, let $\mathcal{C}(p)$ denote the *giant strongly connected component* (GSCC) of the thinned graph (when it exists), i.e. a component occupying a nonvanishing fraction of nodes in the large-$n$ limit. We interpret $\mathcal{C}(p)$ as the *global recycling core*: inside $\mathcal{C}(p)$, imbalances can be re-routed through alternative directed pathways, while outside it, deficits cannot reliably be financed by multilateral chains once key links fail.

Thus, a *topological phase transition* occurs at the critical $p_c$ where the GSCC ceases to exist (or collapses to microscopic size). This is the precise mathematical counterpart of a regime shift from "recycling possible" to "recycling impossible".

## 6.3 MESSAGE PASSING FOR SPARSE DIRECTED NETWORKS (LOCALLY TREE-LIKE REGIME)

Assume the directed network is sparse and locally tree-like (few short directed cycles compared to size), so that distinct directed neighborhoods are asymptotically independent. Message passing (cavity) methods then provide *exact* percolation equations in the large-$n$ limit.

For each directed edge $e = (i \to j) \in E^{\to}$ define the message

$$u_{i\to j} \in [0,1]$$

to be the probability that, *conditioning on $e$ being present*, following $e$ does *not* lead from $j$ into the giant percolating structure via further occupied edges (in the locally tree-like approximation). To avoid immediate trivial reversals, we exclude the backtracking step $j \to i$ when it exists. Then the self-consistency equations take the form

$$u_{i\to j} \;=\; 1-p \;+\; p \prod_{\substack{(j\to k)\in E^{\to} \\ k\neq i}} u_{j\to k}. \tag{40}$$

The intuition is: either $e$ is absent (probability $1-p$), or it is present (probability $p$) and *all* forward non-backtracking continuations from $j$ fail to reach the giant structure.

Given the fixed point $\{u_{i\to j}\}$, node-level percolation probabilities follow. For example, the probability that node $i$ does *not* connect to the giant "out" structure is

$$U_i^{\text{out}} \;=\; \prod_{(i\to k)\in E^{\to}} u_{i\to k}, \tag{41}$$

so that $1 - U_i^{\text{out}}$ is the probability that $i$ reaches the giant out-component. Analogous "in" messages may be defined on the reversed graph to characterize giant in-components; the GSCC corresponds (heuristically and in standard random-graph limits) to the intersection of giant in- and out-components.

## 6.4 NON-BACKTRACKING OPERATOR AND THE CRITICAL THRESHOLD

Index directed edges by $E^{\to}$ and define the non-backtracking (Hashimoto) matrix $B_{\text{nb}} \in \{0,1\}^{|E^{\to}|\times|E^{\to}|}$ by

$$(B_{\text{nb}})_{(i\to j),(k\to \ell)} \;=\; \begin{cases} 1, & \text{if } j = k \text{ and } \ell \neq i, \\ 0, & \text{otherwise.} \end{cases} \tag{42}$$

This is exactly the adjacency operator of the directed line graph under the constraint of no immediate backtracking. Historically, this operator is central in Ihara zeta function theory and is often called the Hashimoto matrix.

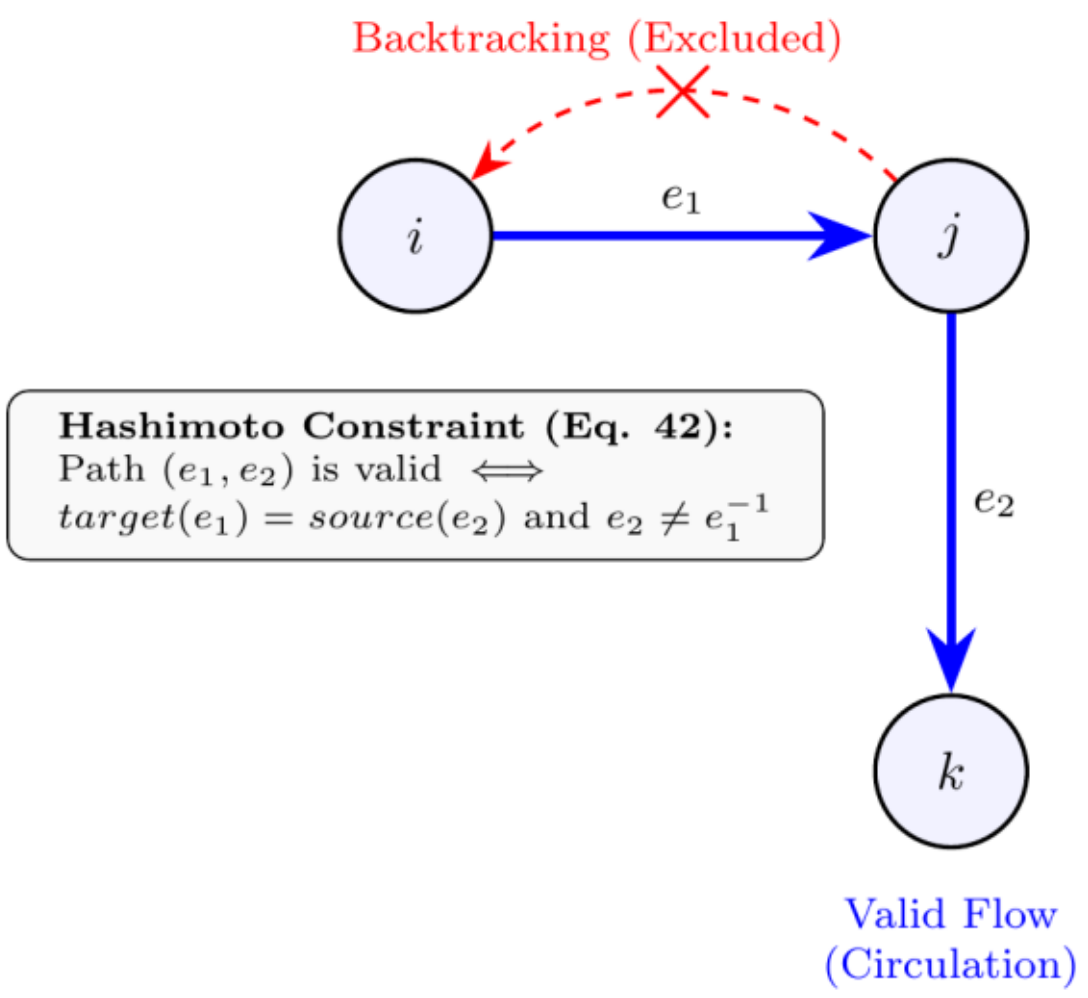

**Figure 1.** The Hashimoto non-backtracking constraint. In Section 6.4, the operator $B_{nb}$ acts on directed edges; a transition from $e_1(i \to j)$ to $e_2(j \to k)$ is permitted, but the immediate reversal back to $i$ is strictly excluded to isolate systemic circulation from bilateral noise.

**Theorem 6.1** (Spectral critical point via linearization of message passing). Assume the directed exposure network is sparse and locally tree-like, so that (40) is exact. Let $\rho(B_{\mathrm{nb}})$ be the spectral radius of $B_{\mathrm{nb}}$. Then the trivial fixed point $u_{i\to j} \equiv 1$ is stable iff $p\,\rho(B_{\mathrm{nb}}) < 1$ and unstable iff $p\,\rho(B_{\mathrm{nb}}) > 1$. Consequently, the bond-percolation threshold is

$$p_c \;=\; \frac{1}{\rho(B_{\mathrm{nb}})}. \tag{43}$$

Proof. Consider the fixed point $u_{i\to j} \equiv 1$, corresponding to the absence of any giant percolating structure. Write $u_{i\to j} = 1 - \epsilon_{i\to j}$ with $\epsilon_{i\to j} \geq 0$ small and expand (40) to first order. Using $\prod_m (1-\epsilon_m) = 1 - \sum_m \epsilon_m + O(\|\epsilon\|^2)$ gives

$$1 - \epsilon_{i\to j} = 1 - p + p\left(1 - \sum_{\substack{(j\to k)\in E^{\to}\\ k\neq i}} \epsilon_{j\to k}\right) + O(\|\epsilon\|^2),$$

hence

$$\epsilon_{i\to j} = p \sum_{\substack{(j\to k)\in E^{\to}\\ k\neq i}} \epsilon_{j\to k} + O(\|\epsilon\|^2). \tag{44}$$

In vector form on the edge-indexed space, (44) is

$$\epsilon = p\,B_{\mathrm{nb}}\,\epsilon + O(\|\epsilon\|^2).$$

Therefore, the linearized dynamics expands perturbations iff $p\,\rho(B_{\mathrm{nb}}) > 1$ and contracts them iff $p\,\rho(B_{\mathrm{nb}}) < 1$. The critical point is the boundary $p_c = 1/\rho(B_{\mathrm{nb}})$. □

Remark 6.1 (Economic meaning of $p_c$ as a phase boundary). The quantity $\rho(B_{\mathrm{nb}})$ is a purely topological branching factor for directed, non-redundant funding/trade chains. When $p > p_c$, the system possesses a macroscopic recycling core and localized imbalances can (in principle) be rerouted around failures. When $p < p_c$, the global exposure graph fragments and the recycling mechanism becomes topologically impossible, generating cascade-like external-adjustment pressure.

## 6.5 HETEROGENEOUS LINK VIABILITY AND A WEIGHTED NON-BACKTRACKING CRITERION

Uniform $p$ is a clean phase diagram; empirically, link viability is heterogeneous. Let $p_{i\to j} \in [0,1]$ be an edge-specific retention probability (e.g. reflecting currency mismatch, maturity structure, sanctions exposure, or bilateral political risk). Define the *weighted* non-backtracking operator $\mathcal{B}$ by

$$\mathcal{B}_{(i\to j),(j\to k)} = p_{j\to k} \quad \text{when } k \neq i, \qquad 0 \text{ otherwise.}$$

Then the same linearization argument yields the generalized criticality condition

$$\rho(\mathcal{B}) \;=\; 1, \tag{45}$$

which reduces to (43) when $p_{i\to j} \equiv p$.

### 6.6 CORE–PERIPHERY LOCALIZATION AND "FALSE STABILITY" OF THE LEADING-EIGENVALUE ESTIMATE

In networks with dense cores embedded in sparse peripheries, the leading eigenvector of $B_{\text{nb}}$ may localize on a small subset of edges. Then $1/\rho(B_{\text{nb}})$ can *underestimate* the true threshold: topology appears stable because a small core has high branching, yet the periphery (where most nodes live) does not percolate. Recent spectral refinements propose searching for the largest *delocalized* real eigenpair of $B_{\text{nb}}$ to better approximate the observed transition in such core–periphery regimes. This is especially relevant for the world economy, whose trade/finance topology is empirically core–periphery.

## 7. TRIFFIN'S DILEMMA AS A NETWORK CONSTRAINT IN A MULTIPOLAR GRAPH

Reserve-currency status can be interpreted in the present framework as an operator perturbation concentrated around one or several central nodes. Let $r \in V$ denote a reserve issuer and consider

$$W(\lambda) = W + \lambda H, \qquad H \geq 0,\ \lambda \geq 0, \tag{46}$$

where $H$ adds exposure mass on channels supported by reserve intermediation. Since $W(\lambda)$ is entrywise increasing, Perron–Frobenius monotonicity implies that the dominant spectral scale cannot fall when reserve-liability supply expands.

**Proposition 7.1** (Reserve supply as a spectral trade-off). *Assume $W$ is irreducible and let $\ell^\top W = \rho(W)\ell^\top$, $Wr = \rho(W)r$, with $\ell^\top r = 1$. Then $\rho\big(W(\lambda)\big)$ is nondecreasing in $\lambda$, and*

$$\frac{d}{d\lambda}\rho\big(W(\lambda)\big)\Big|_{\lambda=0} = \ell^\top H r. \tag{47}$$

*If $H$ loads positively on cycle-supporting edges, the derivative is strictly positive.*

The interpretation is the Triffin tension in operator form: reserve provision can thicken connectivity and settlement viability while simultaneously enlarging the dominant amplification scale. In a multipolar system the same logic applies with several central issuers, and reduced spectral separation between leading modes makes systemic rankings more sensitive to perturbations. Thus reserve architecture is a perturbation problem for the exposure operator.

## 8. POLICY: A GLOBAL CLEARING UNION AS SPECTRAL NETWORK CONTROL

The mathematical policy objective is to keep the system away from two phase boundaries: amplification instability, captured by $\rho(B) \uparrow 1$, and topological fragmentation, captured by $p \downarrow p_c$. This suggests a network-control formulation in which international adjustment rules act not only on observed imbalances, but on the operators that govern propagation and connectivity.

A minimal clearing mechanism uses balances $q_i(t)$ and symmetric adjustment transfers $s_i(t)$,

$$q_i(t+1) = q_i(t) + b_i(t) - s_i(t), \tag{48}$$

with the rule that persistent deficits and surpluses are both penalized. To internalize systemic externalities, one may weight countries by the centrality objects developed above and impose a quadratic loss

$$\mathcal{J}(q) = \sum_{i=1}^{n} w_i\, q_i^2, \tag{49}$$

where $w_i$ increases with Perron/PageRank importance. A stability-preserving clearing design is then summarized by the explicit spectral constraints

$$\rho\big(B(t)\big) \leq \bar{\rho} < 1, \qquad p(t) \geq \bar{p} > p_c(t). \tag{50}$$

These inequalities formalize the idea that a viable global clearing arrangement must simultaneously damp amplification and preserve the recycling core.

At the margin, the derivative formulas from Section 4 identify where intervention is most effective: edges with large $\ell_i r_j$ contribute most strongly to the dominant spectral scale, so taxes, liquidity buffers, collateral rules, or position limits should be concentrated there rather than distributed uniformly. In this sense a clearing union is best understood as spectral network control rather than as a purely accounting device.

## 9. IMPLEMENTATION ROADMAP (EMPIRICAL LAYER, WITHOUT AD HOC MODELING)

The framework is empirically operational once bilateral trade, portfolio, banking, and investment-position data are mapped into a directed exposure matrix. A practical pipeline is: construct a signed bilateral-flow matrix from BoP-consistent sources, pass to the nonnegative exposure lift $W$, compute centralities and marginal spectral impacts, estimate the propagation operator $B$ and its stability margin, and then evaluate diffusion and percolation diagnostics on the induced network.

The empirical aim of the present paper is modest. We do not fit a full macroeconometric model here; instead, we specify a reproducible operator-based measurement layer that can be implemented from standard sources such as IMF BoP/IIP and partner-country datasets together with BIS banking exposures. Once these data are assembled, the proposed quantities are computable by standard linear-algebra and graph-theoretic routines. This keeps the paper within applied-mathematics scope while leaving calibration and forecasting exercises to subsequent work.

## 10. CONCLUSION

Global imbalances are not merely a list of national deficits and surpluses; they form a directed network of interdependent exposures whose stability is governed by operator structure. Once expressed as an adjacency/exposure operator, systemic risk becomes spectral and therefore *computable* from measurement-grade data: the Perron root controls endogenous amplification; eigenvector/PageRank geometry identifies systemically important imbalance placement; the directed Laplacian formalizes diffusion and dissipation of stress; and, under explicit sparse-network assumptions, the percolation critical point is characterized by the leading eigenvalue of the non-backtracking operator.

The resulting research program is simultaneously rigorous and operational: it yields stability metrics, phase boundaries, and marginal systemic-attribution formulas that do not depend on a particular macro model. Policy likewise becomes operational: global clearing can be designed as *spectral network control* that keeps the system away from amplification and fragmentation transitions, replacing ad hoc bargaining with transparent equilibrium constraints checked directly on the evolving global network.

## ACKNOWLEDGEMENTS

The authors thank the Commissioner of Collegiate Education and the Principal of Government College (Autonomous), Rajahmundry, for support and encouragement of academic research.

## FUNDING

The authors declare that no funds, grants, or other support were received during the preparation of this manuscript.

## AUTHOR CONTRIBUTIONS

All authors contributed substantially to the conception and development of the study, the mathematical and economic interpretation of the framework, critical revision of the manuscript, and approval of the final version. Chandrasekhar Gokavarapu prepared and deposited the original arXiv preprint and serves as the corresponding author. All authors accept responsibility for the integrity of the work.

## COMPETING INTERESTS

The authors declare that they have no competing financial or non-financial interests.

## ETHICS STATEMENT

This study uses only publicly available, aggregated country-level macroeconomic and financial statistics. It does not involve human participants, human data, animal subjects, or clinical trials, and therefore does not require ethical approval or informed consent.

## DATA AVAILABILITY

No new empirical dataset was generated or analyzed for the theoretical results in this article. The implementation roadmap identifies publicly available official sources, including IMF Data, BIS statistics, and the United Nations Comtrade database. No proprietary or confidential data underlie the mathematical results.

## PREPRINT STATEMENT AND AUTHORSHIP CONFIRMATION

An earlier version of this work was deposited by Chandrasekhar Gokavarapu on arXiv on 29 January 2026 as *The Spectral Topology of Global Imbalances: A Graph-Theoretic Framework for Systemic Risk in the Balance of Payments* (arXiv:2602.02541; DOI: 10.48550/arXiv.2602.02541). The present manuscript is the revised, coauthored journal version of that work and is not a separate or previously published article. All authors have reviewed and approved the present version, consent to its submission and publication, and confirm the authorship statement above.